\documentclass[a4paper]{amsart}

\title[PDE Models and Spatial Heterogeneity]{Partial differential equation models for invasive species spread in the presence of spatial heterogeneity}

\author[Hughes, Moyers-Gonzalez, Murray, and Wilson]{Elliott Hughes \and Miguel Moyers-Gonzalez \and Rua Murray \and Phillip L.\! Wilson.}
\address{School of Mathematics and Statistics, University of Canterbury}
\email[Corresponding Author]{phillip.wilson@canterbury.ac.nz}

\usepackage{amsmath}
\usepackage{amssymb}
\usepackage{graphicx}
\graphicspath{{./}}
\usepackage{commath}
\usepackage{textcomp}
\usepackage{gensymb}
\usepackage[colorlinks=true,linkcolor=blue,citecolor=blue]{hyperref}
\usepackage{caption}
\usepackage{subcaption}
\usepackage[backend=biber, style=numeric, maxbibnames = 99, giveninits=true]{biblatex}
\DeclareDelimFormat{finalnamedelim}{\addspace\bibstring{and}\space}
\renewbibmacro{in:}{%
  \ifentrytype{article}{}{\printtext{\bibstring{in}\intitlepunct}}}

\usepackage{chngcntr}
\counterwithin{equation}{section}
\usepackage[nameinlink,capitalize]{cleveref}

\usepackage{makecell}
\usepackage{float}

\begin{document}

\begin{abstract}
Models of invasive species spread often assume that landscapes are spatially homogeneous; thus simplifying analysis but potentially reducing accuracy. We extend 
a recently developed partial differential equation model for invasive conifer spread to account for spatial heterogeneity in parameter values and introduce a 
method to obtain key outputs (e.g.\! spread rates) from computational simulations. Simulations produce patterns of spatial spread remarkably similar to 
observed 
patterns in grassland ecosystems invaded by exotic conifers, validating our spatially explicit strategy. We find that incorporating spatial 
variation in different parameters does not significantly affect the evolution of invasions (which are characterised by a long quiescent period followed by 
rapid evolution towards to a constant rate of invasion) but that distributional assumptions can have a significant impact on the spread rate of invasions. Our 
work demonstrates that spatial variation in site-suitability or other parameters can have a significant impact on invasions and must be considered when 
designing models of invasive species spread.
\end{abstract}

\maketitle

\section*{Introduction}
Invasive species pose a significant risk to biodiversity in environments across the globe \cite{caplat2012modeling}. As well as the loss of biodiversity that occurs when introduced organisms 
dominate ecosystems or eliminate native species, the resulting changes to ecosystems can increase fire risk, cause drought and create a range of other hazards (see \cite{MPI_pine_report} for an example). 
Consequently, understanding the processes by which invasive species colonise vulnerable landscapes is an important goal of ecological research. Information on 
how a biological invasion may evolve over time allows managers and conservation practitioners to better allocate resources, prioritise responses to multiple 
threats, and develop strategies to prevent or mitigate the impacts of an invasion. On a societal level, understanding the evolution of a biological invasion increases 
accuracy of estimates of the risks posed by invasions and the costs of inaction. 

Mathematical modelling is frequently employed in ecological research \cite{thompson2021mechanistic,murray_book}. Models of ecological processes have been used to understand the stability 
of ecosystems to perturbations \cite{murray_book}, complex dynamics in the reproductive biology of parasites \cite{schneider2018adaptive}, and (with particular relevance to this work) to understand and 
predict the spread of invasive species \cite{thompson2021mechanistic}. Such models of invasive species spread can provide valuable insights to managers \cite{thompson2021mechanistic} 
and these insights have been employed to successfully manage biological invasions \cite{gosling1989eradication}. For example in \cite{gosling1989eradication} 
models were employed to simulate the response of Coytus (an invasive rodent) populations to climactic variation and management effort, enabling a successful 
eradication program to be implemented. 

Models for invasive species spread can be continuous (e.g.\! \cite{jones2017squirrel}) or discrete in time (e.g.\! \cite{caplat2012seed}) and often feature a spatial component. Such explicitly spatial models 
are particularly valuable in cases where invasive species are still spreading throughout a landscape (or network of landscapes) and thus the size of the invaded 
region may be a particularly important quantity of interest to management specialists. Spatial models can also simulate the consequences of management strategies 
which vary in space. Since actual invasions are often complex phenomena with multiple source populations or invading fronts, understanding how different 
spatial allocations of effort affect long-term dynamics can be an important output of ecological models \cite{caplat2014cross}. 

However, even spatially explicit models of invasion often ignore within-landscape spatial heterogeneity (e.g. local changes in site-suitability). There is some 
evidence that this approach may be appropriate in certain cases \cite{pauchard2016pine} and imposing homogeneity across a landscape significantly simplifies analysis but the potential 
impact of this assumption is not often explored. In this paper, we explore computationally the impact of allowing spatial variation in parameters in an existing 
model for invasive species spread. We find that incorporating spatial heterogeneity in parameters can have a significant impact on the qualitative appearance of 
the solution. In particular, we find that incorporating spatial heterogeneity results in solutions which appear qualitatively similar to observed invasions. 
We also observe that some features of solutions are conserved under spatial heterogeneity 
(such as overall structure of invasions as almost static for a significant period of time before a rapid evolution towards a constant rate of advance). 

Our results emphasise the importance of spatial structure in mediating invasions. We find that both the existence of variation in parameters and the distribution of parameters in space can significantly 
impact observed dynamics. Whether spatial heterogeneity accelerates or retards invasions depends on the distribution of parameters and the period of time 
considered.

\section*{Prior Literature}
For this study, we consider invasions of grassland ecosystems by exotic conifers. Such invasions are well documented in New Zealand, South Africa, and 
Chile (among other Southern Hemisphere countries) \cite{richardson1994pine}. The most common invasive species are trees from the family \emph{Pinaceae}, although other species 
sometimes also pose risks \cite{MPI_pine_report}. Because these species are often important economically as forestry trees, managers and conservation specialists need information 
to contain invasions and assess the risk of new plantings \cite{MPI_pine_report}. Consequently, there is a moderately sized literature mathematically modelling the spread of exotic 
conifers (sometimes informally known as wilding pines). For a review of these models, see \cite{mythesis} or \cite{hughes2023mathematically}. However, most of these models ignore spatial variation 
in parameters, except \cite{caplat2014cross} and \cite{davis2019simulation}. 

In both \cite{caplat2014cross} and \cite{davis2019simulation}, these papers incorporate spatial heterogeneity into a discrete-space, discrete-time, cellular automata model for pine spread. A given domain 
(8km $\times$ 8km in \cite{caplat2014cross}) is divided into square cells of equal size (20m $\times$ 20m in \cite{caplat2014cross}) and pines are `seeded' onto a small subset of 
these cells. Pine spread is managed by semi-deterministic rules (e.g. seeds from a tree are spread into cells according to some probability distribution but 
growth once established is deterministic). Habitat suitability is either generated from existing maps of a particular landscape \cite{davis2019simulation} or generated using a spatially autocorrelated random process. 
In both cases one or more parameters are scaled by a fixed amount in the `bad' habitat to reduce the establishment rate, fecundity or survival rate of trees 
in these less supportive environments.

In \cite{caplat2014cross} the extent of this bad habitat was shown to affect the proportion of the cells invaded, at least at the level of the management unit (e.g. 2km $\times$ 2km subregions 
used to determine the allocation of management resources)  
but \cite{caplat2014cross} do not claim to find an effect at the regional levels. However, it is important to note that \cite{caplat2014cross} base this assessment on the relative 
importance of parameters, not their absolute importance. Since this analysis is also based on simulated invasions across only three different spatial configurations, it 
is difficult to have high confidence in these predictions. In the case of \cite{davis2019simulation}, only one spatial configuration is considered so it is not possible to draw 
inferences on the impact of spatial heterogeneity on model results. 

\section*{Methods}
To explore the impact of spatial heterogeneity on invasions we employed the model of \cite{hughes2023mathematically}. This partial differential equation (PDE) 
model is well-suited to exploring spatial heterogeneity as it is spatially explicit and invasions can be directly simulated, unlike some IDM models \cite{hughes2023mathematically}. 
Furthermore, unlike the cellular automata models discussed above, this model allows growth rates to depend on the total amount of biomass present in a location (rather 
than keeping growth rates constant until the carrying capacity is reached). To include spatial heterogeneity in this model we first extended it to a 2D 
spatial domain. 

\subsection*{Invasions on a 2D Domain}
\begin{figure}[H]
    \includegraphics{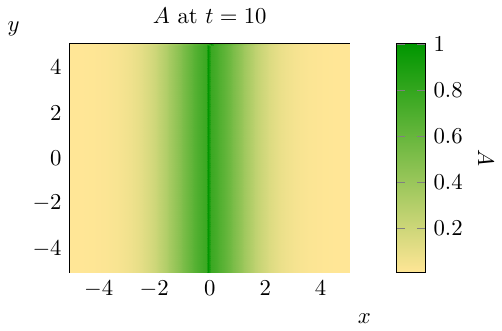}
    \caption{The state of an invasion on a 2D domain at $t = 10$. Parameter values are as in 
    \cite{hughes2023mathematically}.}
    \label{fig:A10_flat}
\end{figure}

In \cite{hughes2023mathematically} a second-order-in-space-first-order-in-time finite differences strategy was used to approximate the model on a one-dimensional domain. 
In a two-dimensional domain, we consider a very simple extension of the model given in \cite{hughes2023mathematically}

\begin{align}
    \frac{\partial A}{\partial t} &= \epsilon C + \rho_0 A - \kappa A^2, \\
    \frac{\partial C}{\partial t} &= D_1\frac{\partial^2 C}{\partial x^2} + D_2\frac{\partial^2 C}{\partial y^2} - C + \frac{A^2}{\beta^2 + A^2},
\end{align}

\noindent 
where $D_1,D_2 \in \mathbb{R}$. Furthermore we impose that spread is approximately isotropic and that $D_1$, $D_2$  
are $\mathcal{O}(1)$ (in practice a correct choice of scales will guarantee that this is the case). In particular we require that $D_1 = D_2 = D \sim 1$ and assume other parameters are of the same order as in 
\cite{hughes2023mathematically}. 
For an initial condition we consider (by analogy with a shelterbelt)

\begin{align}
    A(0,x,y) &= \begin{cases}1 & x \in [-L/n,L/n], \\ 0 & \text{otherwise}, \end{cases} \\
    C(0,x,y) &= 0, 
\end{align}

\noindent 
with Neumann boundary conditions as in \cite{hughes2023mathematically}. To simulate this model using a finite difference method, we apply a second-order 
in space (utilising a standard five-point stencil) first-order in time approach and use ghost points to account 
for the Neumann boundary conditions (see \autoref{fig:A10_flat}).

\subsection*{Spatial Heterogeneity}
With this 2D model in hand, we consider spatial variation in parameter values. 
There are plausible reasons to expect spatial heterogeneity to either increase or 
reduce invasion speeds. One possibility, discussed in \cite{caplat2012modeling}, is that 
spatial heterogeneity will lead to landscape `fragmentation' with invasions slowed or even halted 
by locations with very low suitability for the invasive species blocking onwards spread. Alternatively, one could 
imagine a scenario where the presence of locations with very high suitability led to clusters 
of high population density, driving rapid invasion of the surrounding less suitable areas. The second 
possibility is especially salient in the context of pine trees, where we might expect that long distance dispersal could 
carry seed from high suitability areas over a large portion of the landscape.

As an example of our approach let us first consider a simple extension of our model, allowing initial 
growth rates (e.g.\! $\rho_0$) to vary spatially in a random fashion. This approach models a scenario 
where site suitability varies in space, but total fecundity and maximum density remains constant. 
This represents an advantage over models discussed in \cite{caplat2012modeling} which do not 
disentangle the effects of varying site suitability and fecundity at each location. 

Since we do not fit the model to real world data, we limit ourselves to computational experiments to explore what dynamics are possible. Our new model now becomes 

\begin{align}
    \frac{\partial A}{\partial t} &= \epsilon C + \rho_0(x,y) A(1 - A), \\
    \frac{\partial C}{\partial t} &= D\nabla^2 C - C + \frac{A^2}{\beta^2 + A^2}.
\end{align}
Where we now define $C$ and $A$ as functions from $\mathbb{R}^2 \rightarrow \mathbb{R}$ and $\rho_0$ 
represents a realisation of a random variable. The two-dimensional case was considered here as we expect that 
considering only a one-dimensional representation of the landscape might artificially increase 
habitat fragmentation.

\subsection*{Generation of $\rho_0(x,y)$}\label{subsec:rho_0_gen}
\begin{figure}[H]
    \centering 
    \includegraphics{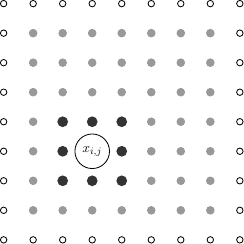}
    \caption{Random variates are created to form a grid of size $N+2\times N+2$ (white, hollow dots). At each of the 
    $N \times N$ inner grey dots $\rho_0$ is calculated by averaging the value of $\alpha$ across the nine neighbouring points 
    on the lattice. At some $x_{i,j} = (x_i,y_j)$ the neighbouring points are shown larger and in dark grey.}
    \label{fig:rho_0_gen}
\end{figure}

We take $\rho_0(x,y)$ at each point on the grid as samples from a 
random variable. In particular, for a grid of size $N \times N$, a grid of size $(N+2)\times(N+2)$ 
of random variates $\alpha_{i,j}$ drawn from a normal distribution with mean one and standard deviation one was generated. Then, for the point $(x_{i},y_{j})$ 
on the lattice of size $N \times N$ the value of $\rho_0(x_{i},y_{j})$ was taken as 

\begin{equation}
    \rho_0(x_{i},y_{j}) = \frac{1}{9}\sum_{k = i-1}^{i+1}\sum_{l =j-1}^{j+1}\alpha_{k,l}.
\end{equation}
The above procedure is best visualised as in \autoref{fig:rho_0_gen}.

\begin{figure}[H]
    \centering 
    \includegraphics{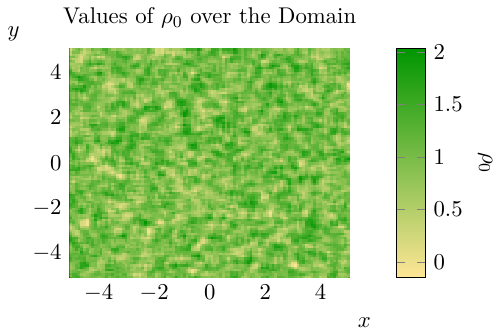}
    \caption{A typical resolution of the random field.}
    \label{fig:rho_0_output}
\end{figure}

Following this procedure implies that the value of $\rho_0$ at some point on the lattice 
is a random variable with mean $1$ and standard deviation $1/3$. Furthermore each variable is 
correlated with its neighbours, such that the lattice varies somewhat smoothly with $x$ and $y$ 
(see \autoref{fig:rho_0_output} for an example of a typical resolution). Note that these parameter values are arbitrary, but are chosen on the theory that 
landscape heterogeneity should be largely continuous and that $\rho_0$ should be positive almost everywhere.

\subsection*{Simulation of an Invasion with Random $\rho_0$}
\begin{figure}[H]
    \centering 
    \includegraphics{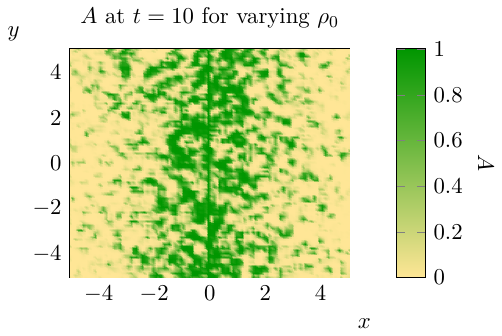}
    \caption{The state of $A$ at $t=10$ for the values of $\rho_0$ given in \autoref{fig:rho_0_output}.}
    \label{fig:example_realization}
\end{figure}

Note that we have allowed $\rho_0$ to be negative, at least for isolated locations on the lattice. Since 
$\rho_0$ is the linearised rate of growth for adult biomass, we will interpret this scenario as 
corresponding to the case where a location is not suitable for pine trees (e.g.\! is rocky or contains a small 
body of water). While we might be concerned about $A(t,x,y)$ becoming negative in such locations, as long as 
$C(t,x,y) \geq 0$ everywhere we are guaranteed that when $A(t,x,y) = 0$ then $\frac{\partial A}{\partial t} \geq 0$. Using the finite difference approximation 
discussed in the previous section, it is relatively straightforward to simulate an invasion for a 
given realisation of $\rho_0$. Since spatial variation occurs only in the $A$ equation, we can write the evolution of the $C$ part of this system as the PDE

\begin{equation}
  \frac{\partial C}{\partial t} = D\left(\frac{\partial^2 C}{\partial x^2} + \frac{\partial^2 C}{\partial y^2}\right) - C + f(t,C,x),
\end{equation}
where $f$ is a strictly bounded function. Consequently it is straightforward to show this method should be stable for sufficiently small $\Delta_t$ 
through a von Neumann analysis (given the stability of $C$ the stability of $A$ follows directly). Given this and an initial condition equivalent to that used for constant $\rho_0$ \autoref{fig:example_realization} shows the 
state of an invasion at $t=10$ over the `landscape' shown in \autoref{fig:rho_0_output}.
One can compare the above solution of the equation with randomly varying $\rho_0$ to the case where $\rho_0$ is 
set at the mean value across the entire domain (see \autoref{fig:A10_flat}). Unsurprisingly there are significant differences between the outcomes of the proceeding two simulations. While there are a number of different dimensions along 
which one could compare these two simulations, we will focus on understanding how spatial heterogeneity changes the spread rate of pine trees. Of course, this 
requires us to define a notion of `spread rate' that can be measured in our computational simulations. 

\subsection*{Spread Rates}
\begin{figure}[H]
    \centering 
    \includegraphics{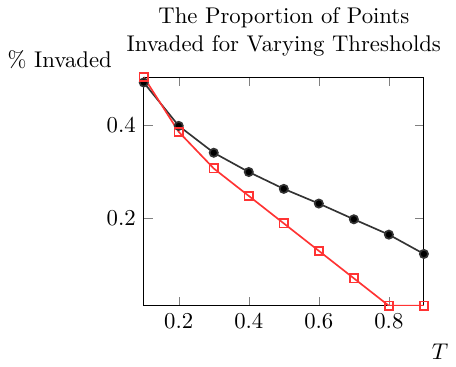}
    \caption{The proportion of points $(x_i,y_j)$ on the lattice with $A(x_i,y_j,10) > T$ for varying $T$, with results for a constant value of $\rho_0$ (red squares) and results for randomly generated values of $\rho_0$ (black circles).}
    \label{fig:pct_invaded}
\end{figure}

We will consider a point on the lattice $(x_i,y_j)$ `invaded' at $t = 10$ if $A(x_i,y_j) > T$, 
where $T$ is some threshold between zero and one. One can now consider the percent of cells on the lattice that are invaded for various thresholds 
for a constant value of $\rho_0$ and for a realisation of our random field (see \autoref{fig:pct_invaded}).

As we can see, at least for this single realisation of the random distribution of $\rho_0$, for the thresholds considered the proportion of invaded cells in 
the random case is equal to or higher than the proportion in the uniform case (except at $T = 0.1$ where the 
proportion invaded is slightly lower). Given this evidence that 
our results are unlikely to be significantly different for different thresholds greater than 0.1, for the rest of this section we will consider a fixed 
threshold $T = 0.5$. In \cite{caplat2012modeling} the authors also consider the proportion of cells or points invaded as a measure of invasion success, but 
since they do not report the criteria they use to determine whether a cell is invaded we cannot compare our approach to that of \cite{caplat2012modeling}. 

\begin{figure}[H]
  \centering
  \hspace*{-0.5in}
  \includegraphics{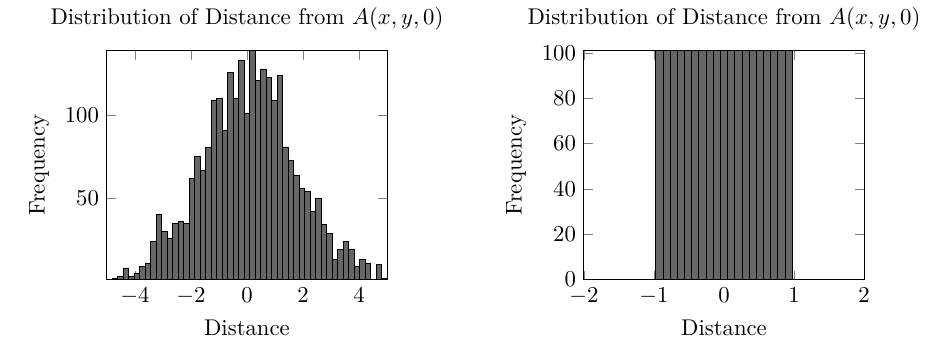}
  \caption{Histogram of the distribution of final signed distances for varying $\rho_0$ (left). Histogram of the distribution of final signed distances for constant $\rho_0$ (right).}
  \label{fig:dist_cells}
\end{figure}

At some time $t_*$ one can consider the distribution of the signed distance of invaded cells to the initial condition $\pm\|(x_i,y_j) - \text{proj}_{A(x,y,0)}(x_i,y_j)\|_2 = x_i$. 
For $t_* = 10$ one can obtain the following histograms for varying $\rho_0$ and constant $\rho_0 = 1$ (see \autoref{fig:dist_cells}).

\begin{figure}[H]
    \centering 
    \includegraphics{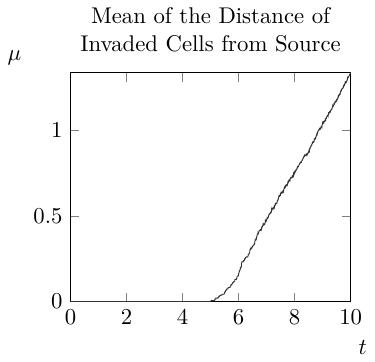}
    \caption{Evolution of the mean distance from the origin over an invasion.}
    \label{fig:mean_list}
\end{figure}

As well as taking snapshots of the distribution for fixed $t$, one can also consider the evolution of the moments of the distribution of unsigned distances as $t$ increases (we 
move from the signed distance to the unsigned distance so that the value of the first moment is not identically zero for all $t$). In 
particular, we consider the evolution of the first two moments of this distribution for varying and constant $\rho_0$. First, consider the evolution of the 
mean of the distribution for varying $\rho_0$ (\autoref{fig:mean_list}). 

\begin{figure}[H]
    \centering 
    \includegraphics{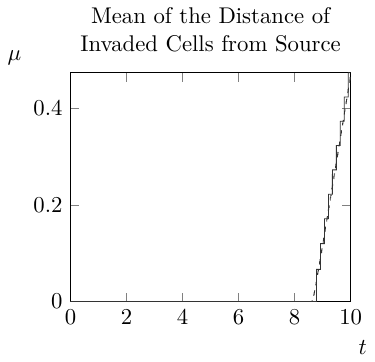}
    \caption{Evolution of the mean of the distance from the origin over an invasion for a constant $\rho_0 = 1$ with $\Delta_x = 10^{-1}$ (solid line) 
    and $\Delta_x = 10^{-2}$ (dashed line).}
    \label{fig:mean_list_flat}
\end{figure}

One can also compute the evolution of the mean for the distribution corresponding to the case with constant $\rho_0$ (see \autoref{fig:mean_list_flat}).
Note that the stepwise nature of the values for $\mu$ from our computational results arise from the discretisation of our domain. If we instead set 
$\Delta_x = 10^{-2}$ (and choose $\Delta_t$ such that the method remains stable) the resulting curve of values for $\mu$ is 
now much smoother (see \autoref{fig:mean_list_flat}).

As can be seen, when $\rho_0$ varies randomly the total proportion of the domain invaded at $t=10$ is higher than for a constant value of $\rho_0$. 
Furthermore, the average distance of invaded cells from the initial condition and the variance of the distribution of distances was much higher when $\rho_0$ 
varied randomly. Thus, in this particular realisation of $\rho_0$, we can see that allowing $\rho_0$ to vary increases the spread of adult pine trees (at least 
at $t=10$).

\subsection*{Inferring Spatial Spread Rates}
\begin{figure}[H]
    \includegraphics{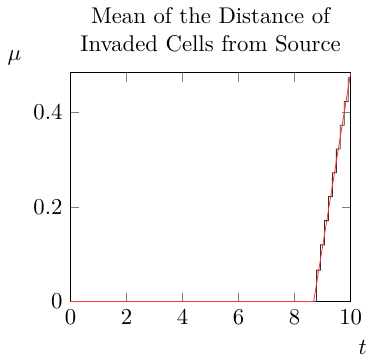}
    \caption{The mean of distance of `invaded' points from the initial condition (black) and a linear fit of the mean distance (red).}
    \label{fig:mean_list_fit}
\end{figure}

One natural question to ask is how the time evolution of the moments of these distributions are related to the asymptotic spread rates that have been computed 
for previous models. Consider first the case where $\rho_0$ is constant across the domain. In this case $A$ and $C$ 
are constant with respect to $y$, so we expect that the solution to our model will depend only on $x$ and $t$. Furthermore, based on previous work we expect that after some critical time $t_*$ the solution for $A$ will approach a travelling wave $f(|\xi|)$, 
$\xi = x - st$ \cite{hughes2023mathematically}. Given this, we expect that for $t > t_*$ we can write the set of `invaded' points (those points such that $A(t,x,y) > T$ for some $T \in [0,1]$) 
as 

\begin{equation}
    \mathcal{I} = [-a - st,a + st] \times [-5,5], a \in \mathbb{R}^+,
\end{equation}
For some $a,s\in \mathbb{R}$. A direct implication is that at time $t$ every point with distance to the initial condition less than $a + st - L/n$ is invaded. Consequently this suggests that the distribution of distances to the 
initial condition at time $t$ should be approximately uniform (except for an isolated peak at $0$, as the initial condition has 
positive measure). However the probability of a randomly selected point landing in the initial condition must be $(2L/n)/(2a + 2st) = L/(n(a + st))$ and thus decreases monotonically as the invasion proceeds. Therefore 
for large $n$ and $t$ we expect the distribution to be approximately uniform and thus have mean 

\begin{equation}
    \mu(t) = \frac{1}{2}(a + st).
\end{equation}
So the average distance of invaded points from the origin moves outwards at a constant speed $s/2$ which is half the speed of the hypothetical 
travelling wave solution. Given this, we fit a piecewise linear function to the mean distance shown in \autoref{fig:mean_list_flat} 

\begin{equation}
    \hat{\mu}(t) = \max \{0, \alpha + \beta t\}.
\end{equation}
Plotting the results (see \autoref{fig:mean_list_fit}), we can see that 
our fitted function closely matches the actual increase in $\mu$ over time.

Since this linear function is a good fit for the data, this suggests that for homogeneous $\rho_0$ and parameter values as in \cite{hughes2023mathematically} the spread speed of invasions is $s \approx 0.72$. To confirm that 
this spread speed was not an artefact of the exact spacing of the spatial discretisation we used, we also simulated an invasion with ten different values of $\Delta_x$ between 
$1/9$ and $1/11$. Then, taking the evolution of the mean distance of invaded points to the initial condition from each `invasion' we fitted our piecewise linear function 
to the resulting $\approx 10^5$ points, obtaining an average spread rate of $s \approx 0.72$. This provides evidence that our estimated spread rate was 
not an artefact of the particular spatial discretisation we used. 

\begin{figure}[H]
    \includegraphics{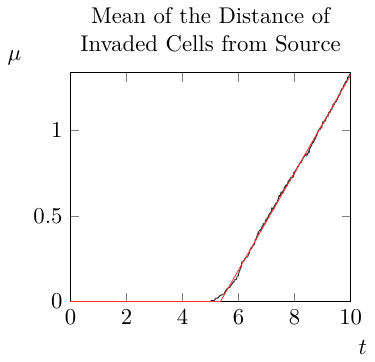}
    \caption{The mean of distance of `invaded' points from the initial condition (black) and a linear fit of the mean distance (red).}
    \label{fig:mean_list_fit_vary}
\end{figure}

Furthermore, this suggests 
a method of inferring spread rates for random grids. To obtain an estimate for the spread rate for randomly generated $\rho_0$ we fit the same piecewise linear function 
to $\mu(t)$ for invasions over the randomly generated `landscape' of varying $\rho_0$ (see \autoref{fig:mean_list_fit_vary}). 

In this particular case we have $s = 0.62$. This suggests that in this case, the spread for this particular set of randomly generated values of $\rho_0$ 
is slightly less than in the case of constant of $\rho_0$. However, since spread occurs much earlier in the case where $\rho_0$ varies the mean distance is 
much higher when $t = 10$ than the case when $\rho_0$ is constant.

\begin{figure}[H]
    \includegraphics{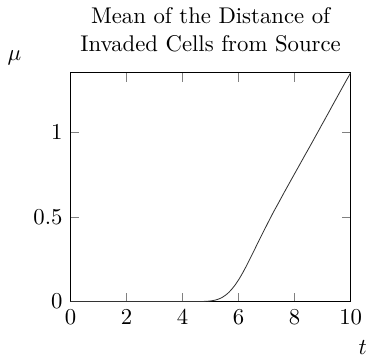}
    \caption{Averaged evolution of the mean distance of invaded cells to the initial condition across one thousand iterations.}
    \label{fig:mean_mean}
\end{figure}

Of course, these results could be an outlier and not reflect the average outcome over a larger number of realisations of the random variable $\rho_0(x_i,y_j)$. 
To consider this possibility we repeated one thousand iterations of our model from $t=0$ to $t=10$, computing at each timestep the average distance of invaded 
points to the initial condition. Then, for each point in time we averaged across all iterations to obtain an average-average distance to the initial condition 
curve (see \autoref{fig:mean_mean}). 

We then construct a piecewise linear fit to obtain an estimate for the average spread rate as above. In this case $s \approx 0.60$, suggesting  
that the spread rate for randomly generated $\rho_0$ will be slower than for constant $\rho_0$. However, because `take off' appears to occur 
earlier when $\rho_0$ is generated randomly, the mean distance of invaded points at $t=10$ is larger for randomly generated $\rho_0$ than for constant $\rho_0$. 
Given that introducing random spatial variation in at least one parameter in the system can have a significant impact on the 
dynamics on the system, we might also naturally wonder whether spatial variation in any of the other parameters produces equally important changes in dynamics. 

In particular, as well as $\rho_0$ (which governs growth of populations after establishment) we will also consider changes in seed production and in the carrying 
capacity of environments. Given that in the previous section we allow the growth rate of trees to vary but keep carrying capacity fixed, it is natural to investigate 
how holding the growth rates constant while allowing carrying capacity fixed may impact spread. In ecological terms, varying $\rho_0$ could represent a situation 
where resource quality varies across a landscape, while varying carrying capacity varies resource availability. The fecundity of adult trees is a natural factor 
to vary, given that variation in seed production is well known to vary significantly between trees \cite{coutts2012reproductive}.

Variation in the population growth (e.g.\! $\rho_0$) is easy to introduce (see the previous section). To vary the carrying capacity we must first note that, in our model, the 
carrying capacity $K_*$ of a particular location is the equilibrium amount of adult biomass in that location, or 

\begin{equation}
    K_* = 1 + \mathcal{O}(\epsilon).
\end{equation}
To allow for spatial variation, we shall reintroduce the parameter $\kappa$, so that the equation for $A$ becomes 

\begin{equation}
    \frac{\partial A}{\partial t} = \epsilon C + \rho_0 A - \kappa A^2,
\end{equation}
Where $\epsilon \sim 10^{-2}$ and $\rho_0, \kappa, C \sim 1$. The carrying capacity now becomes $K_* = \rho_0/\kappa + \mathcal{O}(\epsilon)$ 
so to impose spatial variation in $K_*$ without affecting the establishment rate of adult trees, we shall allow $\kappa$ to vary spatially. Finally, to 
vary seed production by location we will reintroduce the parameter $\gamma$ and allow it to vary spatially (that is, we will allow seed production to 
vary around the base rate by which we scaled $C$ in the previous chapter).

In \cite{mythesis} it has been demonstrated that different distributional assumptions on parameters can significantly impact outcomes (at least in some models), we will compare the impact of allowing each of 
the parameters to vary under three different distributional assumptions. Before proceeding further we will establish a little notation. Firstly, let us denote the spatial averaging operation discussed in \autoref{subsec:rho_0_gen} 
as $S$, so that the spatially averaged random variable discussed in the previous section is $S(Z)$ (where $Z$ is a normally distributed random variable with mean one and standard deviation one). Then we 
will consider the following nine possible combinations of variables and distributions.

\begin{table}[H]
    \centering
\begin{tabular}{|c|c|c|c|}
    \hline
     & $X_1 = \text{truncnorm}(0.1,1.9,1,0.3)$ & $X_2 = \text{uniform}(0.1,1.9)$ & $X_3 = \exp(0.9) + 0.1$ \\
    \hline
    $\rho_0$ & $\rho_0 \sim S(X_1)$ & $\rho_0 \sim S(X_2)$ & $\rho_0 \sim S(X_3)$ \\
    \hline
    $\kappa$ & $\kappa \sim S(X_1)$ & $\kappa \sim S(X_2)$ & $\kappa \sim S(X_3)$ \\
    \hline
    $\gamma$ & $\gamma \sim S(X_1)$ & $\gamma \sim S(X_2)$ & $\gamma \sim S(X_3)$ \\
    \hline
\end{tabular}
\caption{Combinations of distributions and parameters considered (note that $\text{truncnorm}(a,b,c,d)$ refers to a truncated normal distribution with 
bounds $a$ and $b$ and with $\mu = c$, $\sigma^2 = d$).}
\end{table}
Note that all parameters have been chosen so that these distributions are bounded below by 0.1 (since $\kappa$ or $\gamma$ are bounded below at zero by biological 
constraints) and are such that $E(S(X_i)) = 1$ for all $i$. However, we expect that the uniform and exponential distributions will produce more extreme values 
than the truncated normal distribution, with the exponential distribution having a particularly high chance of producing large positive outliers. 

\section*{Results}
For each of these nine combinations, we will compute the spread rate discussed in the previous section and the average percentage of points invaded for different thresholds 
at $t=10$. Beginning first with $\rho_0 \sim S(X_1)$ we consider each of these nine combinations in turn (see overleaf for \Cref{fig:rho_stats,fig:rho_states,fig:gamma_stats,fig:gamma_states,fig:kappa_stats,fig:kappa_states}).

\clearpage 
\begin{figure}[H]
  \includegraphics{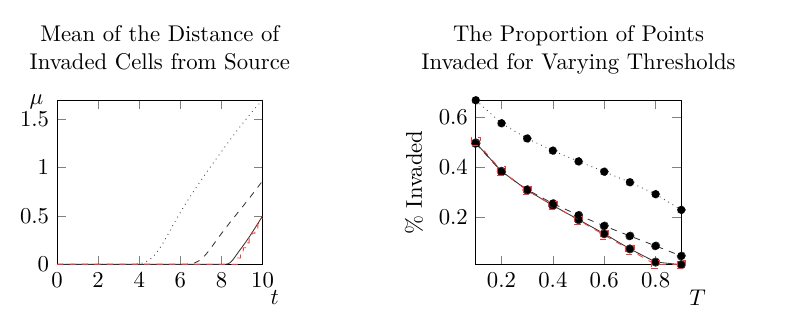}
  \caption{Mean distances from invaded cells to the initial condition over time (left) and proportion of points invaded at $t=10$ for differing thresholds (
    right) for $\rho_0 \sim S(X_1)$ (solid black line), $\rho_0 \sim S(X_2)$ (dashed black line), $\rho_0 \sim S(X_3)$ (dotted black line) and 
    $\gamma = 1$ (dashed red line).}
    \label{fig:rho_stats}
\end{figure}
\begin{figure}[H]
    \centering
    \hspace*{-0.5in}
    \includegraphics{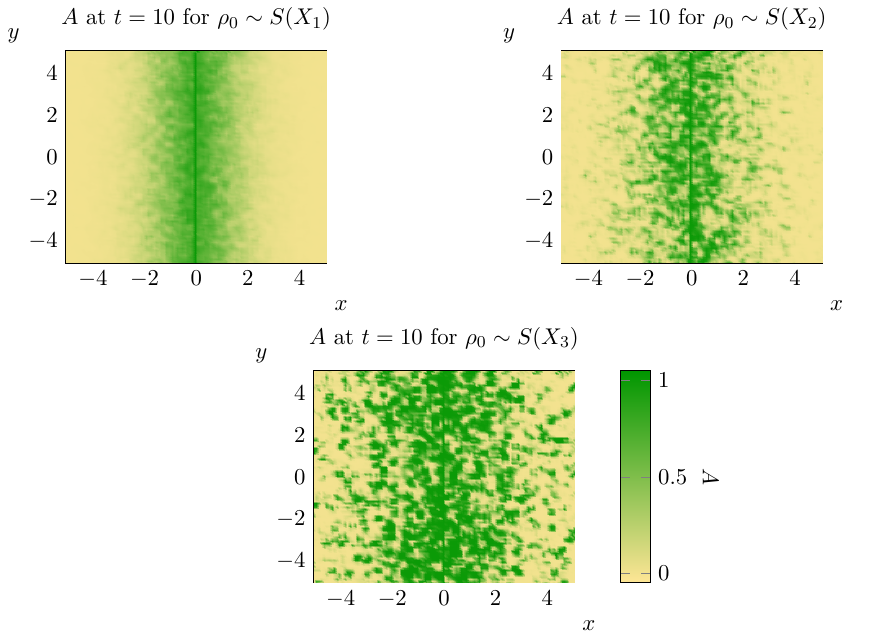}
    \caption{Representative states for $\rho_0 \sim S(X_1)$ (top left), $\rho_0 \sim S(X_2)$ (top right) and $\rho_0 \sim S(X_3)$ (bottom) at $t=10$. See bottom 
    left for the scale for all three states.}
    \label{fig:rho_states}
\end{figure}

\begin{figure}[H]
    \includegraphics{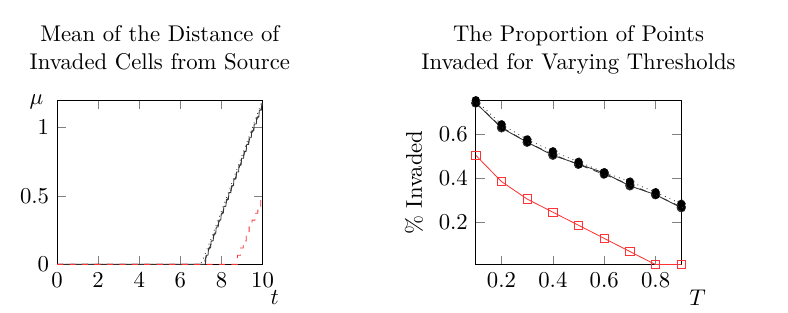}
    \caption{Mean distances from invaded cells to the initial condition over time (left) and proportion of points invaded at $t=10$ for differing thresholds (right) for $\gamma \sim S(X_1)$ (solid black line), $\gamma \sim S(X_2)$ (dashed black line), 
    $\gamma \sim S(X_3)$ (dotted black line) and $\gamma = 1$ (dashed red line).}
    \label{fig:gamma_stats}
\end{figure}
\begin{figure}[H]
    \centering
    \hspace*{-0.5in}
    \includegraphics{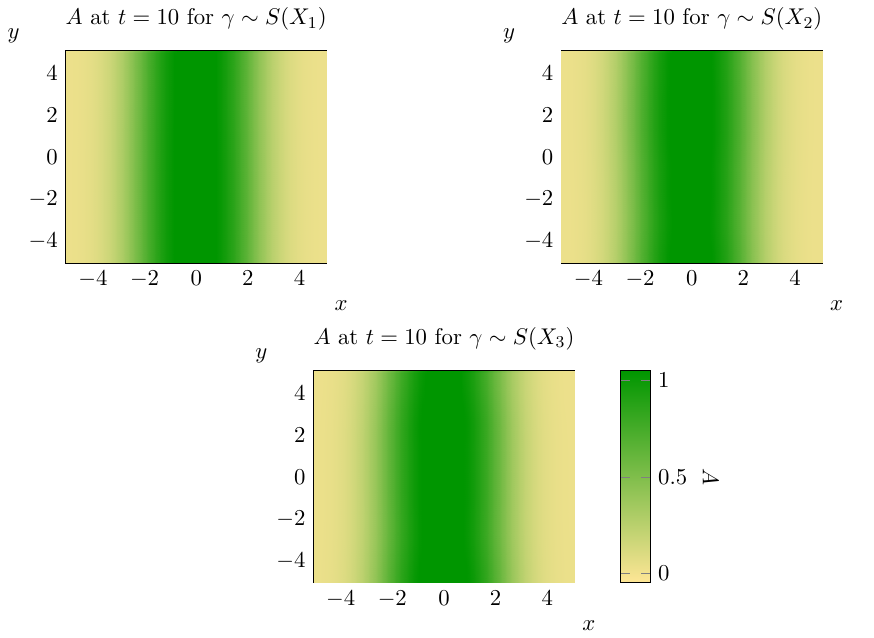}
    \caption{Representative states for $\gamma \sim S(X_1)$ (top left), $\gamma \sim S(X_2)$ (top right) and $\gamma \sim S(X_3)$ (bottom) at $t=10$. See bottom 
    left for the scale for all three states.}
    \label{fig:gamma_states}
\end{figure}

\begin{figure}[H]
    \includegraphics{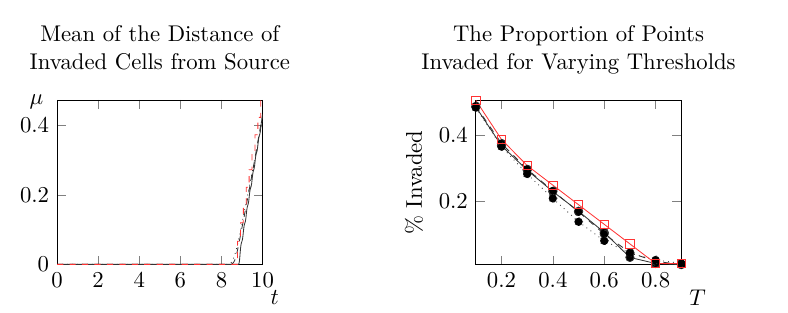}
    \caption{Mean distances from invaded cells to the initial condition over time (left) and proportion of points invaded at $t=10$ for differing thresholds (right) for $\kappa \sim S(X_1)$ (solid black line), $\kappa \sim S(X_2)$ (dashed black line), 
    $\kappa \sim S(X_3)$ (dotted black line) and $\kappa = 1$ (dashed red line).}
    \label{fig:kappa_stats}
\end{figure}
\begin{figure}[H]
    \centering
    \hspace*{-0.5in}
    \includegraphics{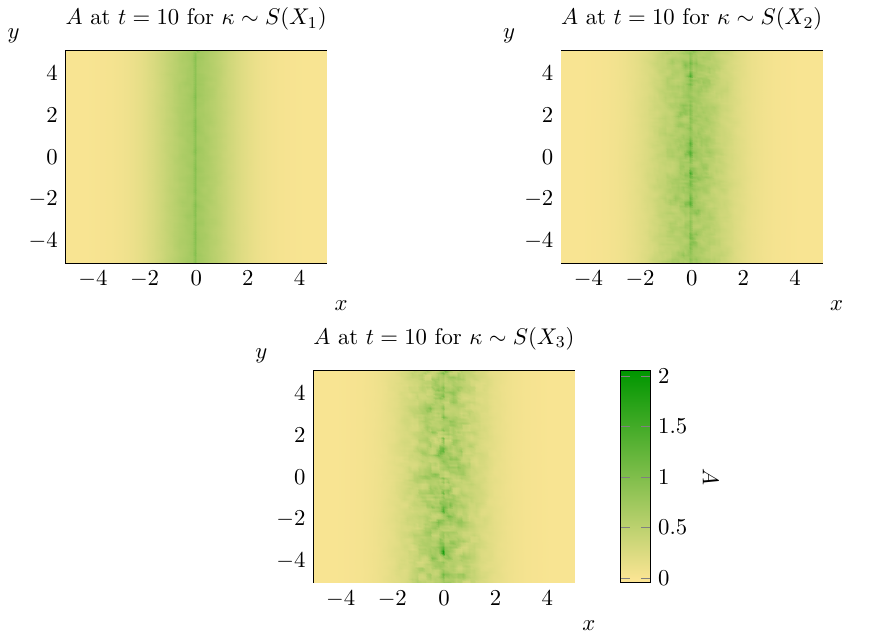}
    \caption{Representative states for $\kappa \sim S(X_1)$ (top left), $\kappa \sim S(X_2)$ (top right) and $\kappa \sim S(X_3)$ (bottom) at $t=10$. See bottom 
    left for the scale for all three states (note that this is different to that of the previous two figures).}
    \label{fig:kappa_states}
\end{figure}

\section*{Discussion}
A few trends are clear from the results of our experiments. Firstly, it is important to note that the impact of spatial variation in a parameter is significantly 
mediated by the combination of parameter and distribution chosen. Spatial variation in $\rho_0$ generally causes spatial spread to begin earlier, but spread rates 
are sometimes lower than with $\rho_0$ constant. Variation in $\gamma$ leads to faster onset of spatial spread and faster spatial spread rates, but changing distributional 
assumptions do not appear to significantly alter outcomes. Notably, the qualitative state of the solution as $t = 10$ is not significantly different from the 
case of constant $\gamma$ under all three distributional assumptions. Changing distributional assumptions for $\kappa$ had a significant impact on the percentage 
of points invaded at $t=10$, but not on the spatial spread rate or the value at which spatial spread began. The spatial spread rates and largest values of $t$ at which $\hat{\mu}(t) =0$ (that is, the time at which spread begins)
are recorded in \autoref{tab:rates} and \autoref{tab:onset} respectively.

\begin{table}[H]
    \centering
\begin{tabular}{|c|c|c|c|}
    \hline
     & $X_1 = \text{truncnorm}(0.1,1.9,1,0.3)$ & $X_2 = \text{uniform}(0.1,1.9)$ & $X_3 = \exp(0.9) + 1$ \\
    \hline
    $\rho_0$ & $0.62$ & $0.54$ & $0.60$ \\
    \hline
    $\kappa$ & $0.72$ & $0.63$ & $0.58$ \\
    \hline
    $\gamma$ & $0.82$ & $0.81$ & $0.81$ \\
    \hline
\end{tabular}
\caption{Spatial spread rates for each of the combinations of parameters and distributions considered (rounded to two decimal places).}
\label{tab:rates}
\end{table}
\begin{table}[H]
    \centering
\begin{tabular}{|c|c|c|c|}
    \hline
     & $X_1 = \text{truncnorm}(0.1,1.9,1,0.3)$ & $X_2 = \text{uniform}(0.1,1.9)$ & $X_3 = \exp(0.9) + 1$ \\
    \hline
    $\rho_0$ & $8.41$ & $6.83$ & $4.31$ \\
    \hline
    $\kappa$ & $8.81$ & $8.86$ & $8.58$ \\
    \hline
    $\gamma$ & $7.13$ & $7.12$ & $7.02$ \\
    \hline
\end{tabular}
\caption{Times of the onset of spatial spread for each of the combinations of parameters and distributions considered (rounded to two decimal places).}
\label{tab:onset}
\end{table}

Despite the significant impacts of spatial variation, all combinations of parameters and distributions continued to display evidence 
of solutions that do not significantly evolve until a critical time $t_*$, before rapidly 
evolving towards solutions which advance at a constant rate (at least with respect to the mean 
distance of invaded locations to the source population). The possible exception to this rule is 
$\rho_0 \sim X_3$, where it appears that the rate at which $\hat{\mu}(t)$ increases is decreasing for $t \sim 10$. However, even in this case 
our results suggest that spread rates are almost constant for $t \sim 1$. This suggests that the model of \cite{hughes2023mathematically} is at 
least qualitatively robust to random variation in parameters. 

\begin{figure}[H]
    \includegraphics{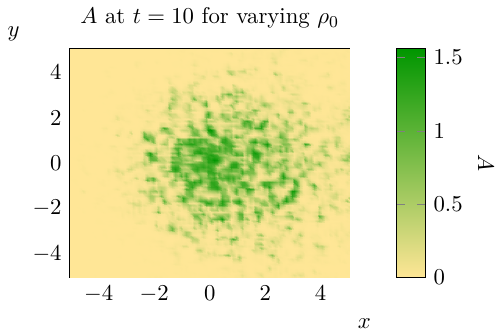}
    \caption{The distribution of $A$ at $t=10$ for a model with an $\mathcal{O}(1)$ advective term and $\rho_0 \sim S(X_1)$. Note that the initial condition 
    in this case is a circular `blob' of width $0.4$ rather than a thin line as in previous plots.}
    \label{fig:fancy_mod}
\end{figure}

Furthermore, at least in some cases, the solutions obtained appear to generate clusters of population growth like those 
observed in invaded landscapes (compare \autoref{fig:rho_states} to Figure 1 in \cite{caplat2012seed} for example). Furthermore, spatial variation in one 
or more parameters can be combined with other extensions to this model to produce spread patterns that are even closer to those observed in invaded landscapes. 
If one considers the advective model advanced in \cite{hughes2023mathematically} but with spatial variation in $\rho_0$ this leads to the following equation 

\begin{align}
  \frac{\partial A}{\partial t} &= \epsilon C + \rho_0(x,y) A(1 - A), \\
  \frac{\partial C}{\partial t} &= D\nabla^2 C + \nu \cdot \nabla C - C + \frac{A^2}{\beta^2 + A^2},
\end{align}
and one can produce patterns which are semi-random and exhibit a preferred 
direction of spread (see \autoref{fig:fancy_mod}).

\section*{Conclusion}
Many models of invasive species spread are deterministic, while the actual invasion process is stochastic. For example, seed production appears to be approximately negative binomial distributed 
among invasive trees \cite{coutts2012reproductive}, so in practice $\gamma$ may vary stochastically between different locations in the landscape. While there 
is some evidence from similar environments overseas that heterogeneity in site suitability may be overwhelmed by the dispersal process \cite{pauchard2016pine}, 
this does not guarantee that random variation in $\gamma$ (e.g. seed production) or other parameters do not impact spread. 

To explore this possibility, we considered extensions to a recent model for exotic conifer invasion \cite{hughes2023mathematically} which allowed a parameter to vary randomly. We showed that, under a variety of parameter choices and 
distributional assumptions, solutions continued to exhibit a constant spread rate (measured by the evolution of the average distance between a unit of biomass and 
the initial condition). Qualitatively and quantitatively, there were significant differences between solutions depending upon the assumptions made, with 
differences in both the asymptotic spread rate and the time at which spatial spread began. Some choices of parameters led to significant increases in the spread 
rate but produced solutions that appeared qualitatively similar to solutions arising from parameters which are constant everywhere. Other solutions produced 
similar spread rates but had a very different qualitative appearance. However, the robustness of the two-regime structure of the solution is promising, as it 
suggests that this model may be resilient to significant local variation in parameter values.

Furthermore, in some cases we observe qualitative patterns that match those observed in New Zealand environments (e.g. highly clustered spread). This provides 
some evidence that a suitable version of our model may produce plausible qualitative pictures of invasions as well as estimates of spread rates or other quantities. This suggests that understanding the spatial patterns in parameters may improve our ability to accurately forecast spread. In some cases, it may be possible 
(e.g.\! by using homogenisation techniques) to `average out' random variation and obtain accurate estimates. In other cases it may be possible to obtain reasonably 
accurate models by neglecting spatial variation in some parameters if the variation does not have a significant impact on the solutions. Alternatively it may 
sometimes be necessary to explicitly include stochastic variation in parameters, although if many parameters are set to vary randomly an agent-based model may 
be a more appropriate choice. In any case, our modelling suggests that spatial heterogeneity may have an important impact on spread rates of pine trees and that 
future research should not ignore the possible impacts of random variation.

It also should be noted that, while we focused on spatial variation induced by random variation in the environment, other sources of spatial variation could 
also be incorporated into our model. For example, future work could explore the impact of different management strategies on spread (as in \cite{caplat2014cross}) 
or environmental phenomena that also evolves over time (e.g.\! coinvasion by herbivores). These time-varying phenomena could potentially have significant impacts on 
the dynamics of solutions and insights from such models may have particularly valuable management implications so future work in this area could be productive.

{ \hypersetup{hidelinks} \printbibliography}

\end{document}